\documentclass[onecolumn,prl,amsmath,amssymb,superscriptaddress,floats]{revtex4-1}
\usepackage[utf8]{inputenc}
\usepackage{amsfonts,amsmath,amssymb}
\usepackage{graphicx}
\usepackage{bm} 
\usepackage{mathrsfs}
\usepackage{subfigure}
\usepackage{textcomp}
\usepackage{hyperref}
\usepackage[flushleft]{threeparttable}
\usepackage[per-mode=symbol]{siunitx}
\usepackage[version=3]{mhchem}

\graphicspath{{../figures/}}

\DeclareSIUnit\intensity{\watt\per\centi\meter\squared}
\DeclareSIUnit\fieldstrength{\volt\per\centi\meter}
\usepackage{xspace}

\DeclareUnicodeCharacter{2009}{\,}

\newcommand{\cost}{\ensuremath{\langle\cos^2\theta_\text{2D}\rangle}}

\newlength{\figwidth}
\newlength{\figwidthwide}
\let\orgautoref\autoref

\renewcommand{\autoref}{%
  \def\equationautorefname{Eq.}%
  \def\figureautorefname{Fig.}%
  \def\subfigureautorefname{Fig.}%
  \orgautoref}

\usepackage{color}
\usepackage[normalem]{ulem}
\definecolor{darkgreen}{rgb}{0.0,0.7,0.0}

\begin{document}


\title{Rotational coherence spectroscopy of molecules in helium nanodroplets: Reconciling the time and the frequency domains}



\author{Adam S. Chatterley} 
\thanks{These two authors contributed equally}

\author{Lars Christiansen} 
\thanks{These two authors contributed equally}

\affiliation{Department of Chemistry, Aarhus University, Langelandsgade 140, DK-8000 Aarhus C, Denmark}

\author{Constant A. Schouder} 
\affiliation{Department of Chemistry, Aarhus University, Langelandsgade 140, DK-8000 Aarhus C, Denmark}

\author{Anders V. J{\o}rgensen} 
\affiliation{Department of Chemistry, Aarhus University, Langelandsgade 140, DK-8000 Aarhus C, Denmark}

\author{Benjamin Shepperson}
\affiliation{Department of Chemistry, Aarhus University, Langelandsgade 140, DK-8000 Aarhus C, Denmark}

\author{Igor N. Cherepanov}
\affiliation{Institute of Science and Technology Austria, Am Campus 1, 3400 Klosterneuburg, Austria}

\author{Giacomo Bighin}
\affiliation{Institute of Science and Technology Austria, Am Campus 1, 3400 Klosterneuburg, Austria}

\author{Robert E. Zillich}
\affiliation{Institute for Theoretical Physics, Johannes Kepler Universit\"{a}t Linz, Altenbergerstraße 69, A-4040 Linz, Austria}

\author{Mikhail Lemeshko}
\affiliation{Institute of Science and Technology Austria, Am Campus 1, 3400 Klosterneuburg, Austria}

\author{Henrik Stapelfeldt} 
\affiliation{Department of Chemistry, Aarhus University, Langelandsgade 140, DK-8000 Aarhus C, Denmark}

\date{\today}

\begin{abstract}
Alignment of \ce{OCS}, \ce{CS_2} and \ce{I_2} molecules embedded in helium nanodroplets is measured as a function of time following rotational excitation by a non-resonant, comparatively weak ps laser pulse. The distinct peaks in the power spectra, obtained by Fourier analysis, are used to determine the rotational, B, and centrifugal distortion, D, constants. For \ce{OCS}, B and D match the values known from IR spectroscopy. For \ce{CS_2} and \ce{I_2}, they are the first experimental results reported. The alignment dynamics calculated from the gas-phase rotational Schr\"{o}dinger equation, using the experimental in-droplet B and D values, agree in detail with the measurement for all three molecules. The rotational spectroscopy technique for molecules in helium droplets introduced here should apply to a range of molecules and complexes.

\end{abstract}

\pacs{}

\maketitle

One of the unique aspects of helium nanodroplets is that molecules embedded in their interior exhibit IR spectra with rotational fine structure very similar to the case of gas phase molecules. This free-rotation behavior, discovered for \ce{SF_6} molecules~\cite{hartmann_rotationally_1995}, is directly connected to the superfluidity of \ce{^4He} droplets because the low density of states below 5 cm$^{-1}$ makes coupling between the molecular rotation and the phonons weak~\cite{toennies_superfluid_2004,choi_infrared_2006,stienkemeier_spectroscopy_2006}. The helium droplet does, however, influence the rotational structure of the molecule through a non-superfluid fraction of the surrounding helium weakly bonding to and co-rotating with the molecule. The result is an effective rotational constant, which is smaller than that of the isolated molecule. For instance, for the \ce{OCS} molecule the reduction factor is 2.8~\cite{grebenev_rotational_2000}.

With this knowledge in mind, one might expect that if molecules in helium droplets could be set into rotation by a short laser pulse, the rotational motion measured as a function of time should resemble that of the corresponding isolated molecules although slowed due to the larger moment of inertia of the helium-solvated molecules.  Using nonadiabatic laser-induced alignment techniques, adopted from gas phase molecules, such time-resolved measurements have been reported~\cite{pentlehner_impulsive_2013,christiansen_alignment_2015,shepperson_laser-induced_2017}. The experiments showed that fs or ps pulses can indeed induce rotation of molecules in He droplets, leading to transient alignment, i.e. confinement of molecular axes to space-fixed axes~\cite{stapelfeldt_colloquium:_2003,fleischer_molecular_2012}. In the case of \ce{I_2} molecules, a weak recurrence of the initial alignment maximum, shortly after the laser pulse, was observed about 600 ps later and, using the angulon model~\cite{schmidt_rotation_2015}, identified as a rotational revival of the helium dressed molecule~\cite{shepperson_laser-induced_2017}. As such, these studies illustrated some similarities to laser-induced rotation of gas-phase molecules but they did not provide a quantitative understanding of the rotational dynamics measured, nor a clear connection to the rotational energy level structure of molecules in He droplets established by frequency-resolved spectroscopy and quantum calculations~\cite{toennies_superfluid_2004,choi_infrared_2006,stienkemeier_spectroscopy_2006,kwon_quantum_2000}.

Here we explore laser-induced alignment of molecules in helium droplets with the rotational energy kept below the roton energy ($\sim$ 5 cm$^{-1}$) to equal the conditions of frequency-resolved spectroscopy studies and thus test if the free-rotation behaviour is also observable in time-resolved measurements. We show that the time-dependent degree of alignment measured can be accurately described by the solution to the time-dependent Schr\"{o}dinger equation for an isolated molecule exposed to the alignment laser pulse, provided the effective $B$ and $D$ constants are employed, and inhomogeneous broadening of rotational energy levels accounted for. Experimentally, \ce{OCS}, \ce{CS_2} and \ce{I_2} molecules are rotationally excited by a non-resonant, picosecond laser pulse and the ensuing degree of alignment recorded by timed Coulomb explosion. Fourier transformation of the alignment traces recorded reveals well-defined spectral peaks corresponding to the frequencies of a rotational wave packet in a non-rigid linear molecule, and thereby $B$ and $D$ for the three molecules are determined. This demonstrates that rotational coherence spectroscopy~\cite{felker_rotational_1992,riehn_high-resolution_2002} works well for linear molecules in He droplets and we believe it applies broadly to a variety of molecules and complexes. We show that $D$ strongly disperses the alignment traces and that the inhomogeneous broadening causes a gradual decay of the oscillatory structure of the alignment degree. The calculated alignment traces enable unambiguous assignment of half and quarter revivals in the experimental data.

The experimental setup and execution is essentially the same as previously reported~\cite{shepperson_laser-induced_2017,shepperson_strongly_2017}. Helium droplets consisting on average of 6--8,000 He atoms and doped with one \ce{OCS}, \ce{CS_2} or \ce{I_2} molecule, are exposed to a linearly polarized laser pulse centered at 800 nm with a 40 nm (FWHM) bandwidth. Its duration (FWHM) is either 15, 5 or 0.45 ps. The purpose of this alignment pulse is to rotationally excite the molecules via the polarizability interaction. After a time, t, the doped droplets are exposed to a 40-fs probe pulse. Its intensity, \SI{6e14}{\intensity}, is sufficiently high to Coulomb explode the molecules. The emission direction of the fragment ions, \ce{^32S^+} for \ce{OCS} and \ce{CS_2}, and \ce{IHe^+} for \ce{I_2}, are recorded by a 2-dimensional detector and the angle, $\theta_\text{2D}$, between the ion hit on the detector and the polarization of the alignment pulse, located in the detector plane, is determined. Hereby \cost, a standard measure for the degree of alignment~\cite{sondergaard_nonadiabatic_2017}, can be determined, where $<$..$>$ means the average over all ion hits, typically 2500 recorded for 50,000 laser pulses. The time-dependent alignment traces, i.e. \cost(t), are obtained by performing the measurements for a large number of t's.

\begin{figure} 
\includegraphics[width=8.5 cm]{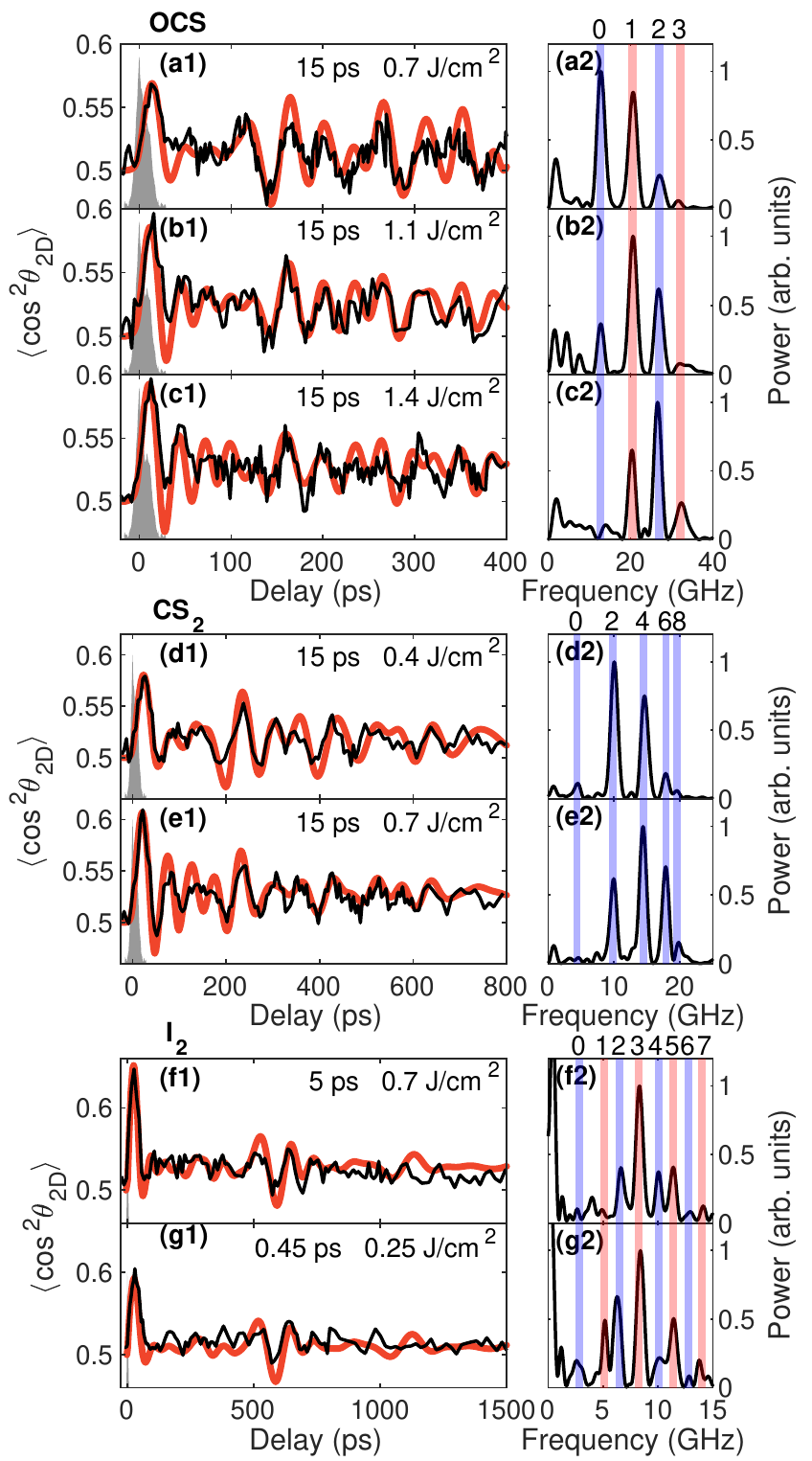}%
\caption{\label{trace-fft} Column 1: The time-dependent degree of alignment for \ce{OCS}, \ce{CS_2} and \ce{I_2} molecules at different durations and fluences of the alignment pulse, given on each panel.  Black (red) curves: experimental (simulated) results. The intensity profile of the alignment pulses are shown by the shaded grey area. Column 2: The power spectra of the corresponding \cost traces. The spectral peaks, highlighted by the colored vertical bands, are assigned as ($J$--$J+2$) coherences (see text) with $J$ given on top of the panels (blue: even, red: odd).}
\end{figure}

The black curves in \autoref{trace-fft}(a1)-(c1) show \cost as a function of time for \ce{OCS}, recorded at three different fluences. In the three cases, a maximum occurring on the trailing edge of the 15 ps alignment pulse is followed by an oscillatory structure. The alignment traces are Fourier transformed to analyze their spectral content. The resulting power spectra, displayed in \autoref{trace-fft}(a2)-(c2) contain distinct peaks, just as for rotational wave packets in gas phase molecules~\cite{sussman_quantum_2006,schroter_crasy:_2011,przystawik_generation_2012,zhang_time-domain_2018}. There, the peaks reflect the frequencies of the nonzero matrix elements {\ensuremath{\langle JM|\cos^2\theta_\text{2D}|J'M\rangle}}, i.e. the coherence (coupling) between state {\ensuremath{|JM\rangle}} and {\ensuremath{|J'M\rangle}} with $J$ denoting the rotational angular momentum and $M$ its projection on the alignment laser polarization~\cite{sondergaard_nonadiabatic_2017}. For a non-rigid linear rotor, the rotational energies are given to the second order by:
\begin{equation}
\label{eq:rotational-energy}
E_\text{rot} = \ensuremath{BJ(J+1)-DJ^2(J+1)^2},
\end{equation}
and thus the frequencies corresponding to the dominant {\ensuremath{\Delta J=J'-J=2}} coherences~\cite{sondergaard_nonadiabatic_2017}, labelled ($J$--$J+2$), by:
\begin{equation}
\label{eq:frequencies}
\nu_\text{(J--J+2)} = B(4J+6) – D(8J^3 + 36J^2 + 60J +36).
\end{equation}
Adopting the gas phase picture, we assign the three prominent peaks in \autoref{trace-fft}(a2) at 12.8, 20.6 and 27.2 GHz as pertaining to the (0--2), (1--3) and (2--4) coherence, respectively. The colored vertical bands illustrate that the peaks are located at the same positions for the three fluences. The weight of the peaks shifts to higher frequencies as the fluence, $F$, increases, and for $F$ = 1.4 J/cm$^2$, an extra peak appears at 32.3 GHz, which we assign as the (3--5) coherence. 

\begin{figure} 
\includegraphics[width=8.5 cm]{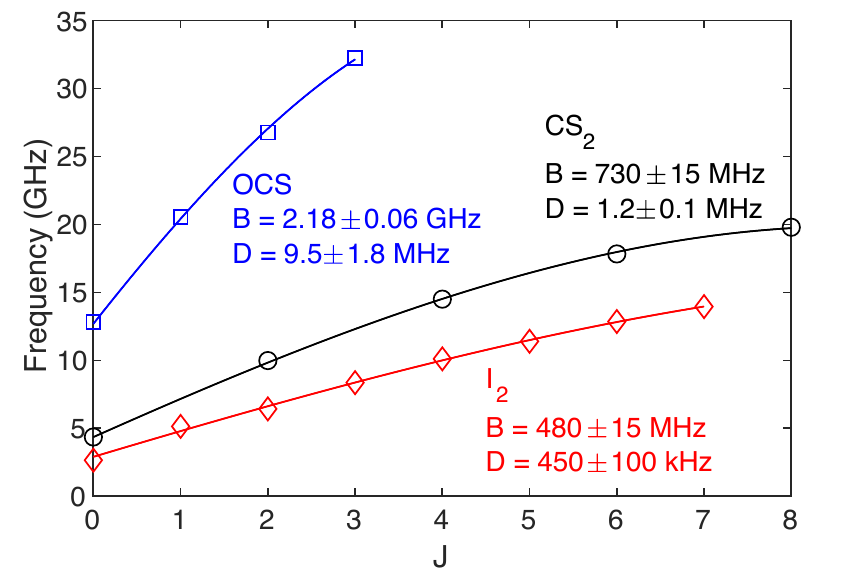}%
\caption{\label{plot-fit} Central frequencies of the ($J$--$J+2$) peaks in the power spectra versus $J$. The full lines represent the best fits using \autoref{eq:frequencies}. The $B$ and $D$ constants from the fits are given for each molecule.  }
\end{figure}

Next the central positions of the ($J$--$J+2$) peaks in \autoref{trace-fft}(a2)-(c2) are plotted as a function of $J$ in \autoref{plot-fit}. The data points, represented by the blue squares, are fitted using \autoref{eq:frequencies} with $B$ and $D$ as the free parameters. The best fit, represented by the blue curve in \autoref{plot-fit}, is obtained for $B$ = 2.18$\pm$0.06 GHz and $D$ = 9.5$\pm$1.8 MHz. These findings agree with the values from IR spectroscopy, $B$ = 2.19 GHz and $D$ = 11.4 MHz, where $D$ was the average for the $v$=0 and $v$=1 vibrational states in the IR transition~\cite{grebenev_rotational_2000}. The excellent fitting of \autoref{eq:frequencies} to the peak positions strongly indicates that laser-induced rotation of \ce{OCS} molecules in He droplets is well described by a gas-phase model employing the effective $B$ and $D$ constants.

To further explore this, we calculated \cost(t) by solving the time-dependent rotational Schr\"{o}dinger equation for a linear molecule exposed to the experimental alignment pulse, using the $B$ and $D$ values from the fit. The calculations were averaged over the initially populated rotational states, given by a Boltzmann distribution with $T$ = 0.37 K, and over the focal volume determined by the measured spot sizes of the alignment ($\omega_\text{0}$ = 30 $\mu$m) and probe beams ($\omega_\text{0}$ = 25 $\mu$m). Also, the effect of inhomogeneous broadening was implemented by a Gaussian distribution of the $B$ constants with a FWHM, $\Delta$$B$ =  90 MHz~\footnote{The width, $W$, of the $J$ = 1--0 rotational transition was measured by microwave spectroscopy to 180 MHz~\cite{lehnig_rotational_2009} which gives  $\Delta$$B$ = $W$$/$2.}, and a constant $B/D$ ratio.

The calculated degree of alignment, shown by the red curves in \autoref{trace-fft}(a1)-(c1), has a very strong resemblance with the measured traces and captures in detail most of the oscillatory pattern observed~\footnote{The simulated data have been scaled by a factor of 0.30, symmetrically centered around \cost = 0.5, to account for the non-axial recoil of the \ce{S^+} ions from the Coulomb explosion of the \ce{OCS} molecules~\cite{shepperson_strongly_2017}}. The good agreement between the experimental and calculated time-dependent \cost~corroborates that an effective gas-phase model accurately describes the laser-induced rotational dynamics of \ce{OCS} in He droplets for the fluences and durations of the alignment pulses employed. Nevertheless, the observed dynamics appears very different from that of isolated molecules. As discussed below, this is due to the much larger $D$ constant for molecules in He droplets compared to isolated molecules (\ce{OCS}: $D_{He} \approx 6.5~\times~10^{3}~D_{gas}$)~\cite{grebenev_rotational_2000,lehmann_rotation_2001} and the presence of inhomogeneous broadening.

Experiments were also conducted on \ce{CS_2} and \ce{I_2} molecules. The alignment traces and corresponding power spectra, including the peak assignments, are shown in \autoref{trace-fft} panels (d)-(g). The central frequencies are plotted versus $J$ and as for \ce{OCS},  \autoref{eq:frequencies} provides excellent fits to the experimental results, illustrated by the red (\ce{I_2}) and blue (\ce{CS_2}) points/lines in \autoref{plot-fit}. The $B$ and $D$ values extracted from the best fits are given on the figure. To our knowledge, this is the first experimental determination of $B$ and $D$ for these two molecules in He droplets. In fact, IR and MW do not apply to \ce{I_2} because it lacks a permanent electric dipole moment.

For comparison, we obtained $B$ for \ce{CS_2} in a helium droplet from a path integral Monte Carlo (PIMC)
simulation, using the \ce{CS_2}-He interaction of Ref.~\cite{yuanTheorChemAcc14} and
the He-He interaction of Ref.~\cite{aziz87}.
The rotational correlation function~\cite{blinovJCP04,zillichJCP05}
$S_\ell(\tau)= {4\pi\over 2\ell +1}{1\over Z}\sum_m {\rm Tr}\{Y^*_{\ell m}(\Omega(\tau))Y_{\ell m}(\Omega(0)) e^{-\beta H}\}$
is calculated in imaginary time $\tau\in[0,\beta]$. $B$ is obtained by fitting
$S_\ell(\tau)$ of a free linear rotor, with $B$ as fit parameter. Unlike a full reconstruction
of $S_\ell(t)$ in real time, the simple fitting procedure is numerically stable for the statistical
errorbars of $S_\ell(\tau)$ achievable for a heavy rotor like \ce{CS_2}. However, the fit is inherently
wrong for $\tau\to 0$ due to sum rules~\cite{zillichJCP05}. We remove this bias by
increasing the imaginary time interval $[0,\beta]$, i.e. decreasing the temperature $T$, and
extrapolate $B$ to $T\to 0$ using a linear fit; we used $T=0.625;0.3125;0.15625$K. Furthermore,
since quantum simulations of droplets of thousands of He atoms are not feasible, we simulated
\ce{CS_2} in clusters of up to $N=150$ He atoms and extrapolated
the results for $B$ to $N\to\infty$ using a fit linear in $N^{-1}$. With this protocol
to remove the bias from the linear rotor fit and from the finite cluster size, the PIMC estimate
is $B_{\rm gas}/B_{\rm He}=4.33\pm0.05$ in large droplets, in good agreement with the experimental value of 4.5$\pm$0.1 ($B_{gas}$ = 3.273 GHz).

For \ce{I_2}, the experiment shows that $B_{He}$ is reduced by a factor of 2.3$\pm$0.1 compared to $B_{gas}$. Previous PIMC simulations for \ce{I_2} in a cluster of 150 He atoms gave a reduction factor of 1.7. From our recent experience with \ce{CS_2}, we expect that the reduction factor will also increase for \ce{I_2} when the new protocol is applied. If we naively assume the relative correction due to the protocol for \ce{I_2} is the same as for \ce{CS_2}, the reduction factor would be 2.3, i.e. the same as the experimental value.

As for \ce{OCS}, we also calculated the time-dependent degree of alignment for \ce{I_2} and \ce{CS_2}. The results, shown by the red curves in Fig. \ref{trace-fft} panels (d1)-(g1) agree very well with the experimental findings~\footnote{The scaling factor of the simulated data to account for non-axial recoil is 0.37 for \ce{CS_2} and 0.75 for \ce{I_2}}. Since the two molecules had not been spectroscopically studied in He droplets before, we had to choose a width for the inhomogeneous distribution of the $B$ constant. The best agreement with the measured \cost(t) was obtained for $\Delta$$B$ = 50 MHz for \ce{CS_2} and 40 MHz for \ce{I_2}.

\begin{figure} 
\includegraphics[width=8.5 cm]{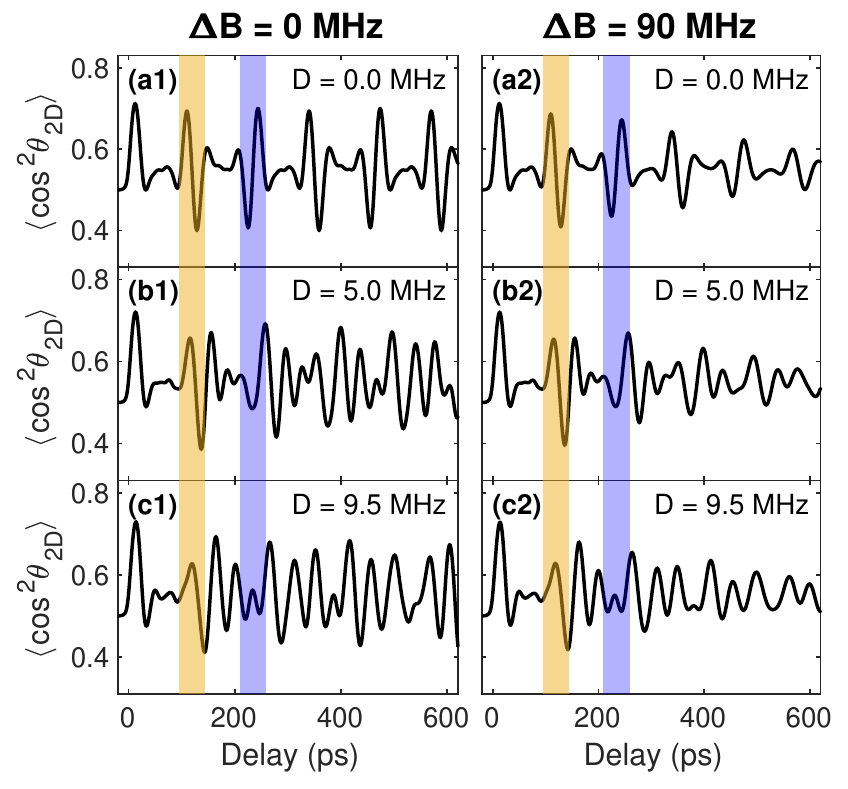}%
\caption{\label{centrifugal-inhomogeneous} \cost~as a function of time calculated for OCS (B=2.17 GHz) for three different values of the D constant, without (left column) or with (right column) inhomogeneous broadening included.  T = 0.37 K and F = 0.7 J/cm$^2$. The yellow and blue bands highlight the position of the half-and full-revival for the D=0 case.}
\end{figure}

To elucidate why laser-induced rotational dynamics of molecules in He droplets appears very different from that of gas phase molecules, we calculated \cost(t) for three values of the $D$ constant, with or without the effect of inhomogeneous broadening. The calculations were done for \ce{OCS} molecules and the experimental 15 ps pulse. When $D$ = 0 and inhomogeneous broadening is neglected, i.e. all molecules have the same $B$ value, \autoref{centrifugal-inhomogeneous}(a1) shows that \cost(t) is periodic with distinct half and full revivals. This case is identical to that of isolated \ce{OCS} molecules except that the revival period is increased by a factor 2.8 due to the effective $B$ constant.  The calculation for $D$ = 5.0 MHz, \autoref{centrifugal-inhomogeneous}(b1), shows that the centrifugal term introduces an additional oscillatory structure in \cost(t)  and distorts the shape of the revivals. For gas phase molecules, it was already observed and understood that the centrifugal term modulates the shape of rotational revivals~\cite{rosenberg_rotational_2017,chatterley_laser-induced_2020} but the influence was moderate~\footnote{Unless extreme rotational states, excited by an optical centrifuge, are populated~\cite{milner_probing_2017}} and the different revivals remained separated from each other. In \autoref{centrifugal-inhomogeneous}(b1), the effect of the centrifugal term is so large that, with the exception of the half-revival, there is essentially no longer distinct, separated revivals. This trend is even more pronounced for the calculation with the experimental $D$ value, \autoref{centrifugal-inhomogeneous}(c1). The yellow and blue bands provide a rigid rotor reference, \autoref{centrifugal-inhomogeneous}(a1), for how the centrifugal term distorts and shifts the rotational revivals.

The panels in the right column of \autoref{centrifugal-inhomogeneous} show \cost(t) when inhomogeneous broadening is included by averaging calculated alignment traces over a 90-MHz-broad Gaussian distribution of $B$ constants. The main influence is a gradual reduction of the amplitude of the oscillations in the alignment traces while preserving the average value of \cost. This dispersion effect produces alignment traces, which (for $D$ = 9.5 MHz) agrees very well with the experimental results for all three fluences studied, see \autoref{trace-fft}(a1)-(c1). The same is true for the \ce{CS_2} results using a 50 MHz distribution of $B$ constants. Calculations where $D$ is gradually increased from 0 to 1.2 MHz (not shown) identify the valley-peak structure around t = 200 ps as the quarter revival, see \autoref{trace-fft}(d1)-(e1). Similarly, for \ce{I_2} the oscillatory structure in the 550-700 ps range \ce{I_2} is identified as the half revival, see \autoref{trace-fft}(f1)-(g1).

We note that the average value of the experimental \cost(t) for \ce{I_2}, recorded with the 5 ps pulse, \autoref{trace-fft}(f1), actually decays over the 1500 ps measured, which is not captured by the calculation. Also, the calculated alignment trace shows a transient structure around t = 1100-1200 ps, corresponding to the full revival, which is not observed experimentally. These two discrepancies between the measurements and simulations indicate that some of the rotational states excited by the alignment pulse decay on the time-scale of the measurements, i.e. that the alignment dynamics is also influenced by lifetime (homogenous) broadening~\cite{choi_infrared_2006,blancafort-jorquera_rotational_2019}. This is likely also the case for \ce{OCS} and \ce{CS_2} but measurement for longer times would have been needed to observe a similar decay of \cost(t). For \ce{I_2}, electric quadrupole induced coupling between the rotational angular momentum and the nuclear spins may also contribute to the decay of \cost(t)~\cite{thomas_hyperfine-structure-induced_2018,yachmenev_laser-induced_2019,thesing_effect_2020}.

The good agreement between the measured and calculated \cost(t) leads us to conclude that, for the relatively weak laser pulses applied here, the mechanism of non-resonant ps or fs laser-induced rotation of molecules in helium droplets is the same as for gas phase molecules. The rotational dynamics does, however, differ significantly from that of isolated molecules due to the orders of magnitudes larger centrifugal constant of He-solvated molecules and the inhomogeneous broadening of the distribution of rotational constants~\cite{lehmann_potential_1999,zillich_lineshape_2008,lehnig_rotational_2009}. Fourier transformation of the measured \cost(t) traces and fitting of the spectral lines to a non-rigid rotor model enabled determination of the rotational and centrifugal constants. For \ce{OCS}, the agreement of our experimental results with the values from IR spectroscopy reconciled frequency resolved spectroscopy and nonadiabatic laser-induced alignment dynamics and in addition, introduced the latter as a rotational spectroscopy method for molecules in helium droplets. Our method should apply to a broad range of molecules~\cite{choi_infrared_2006,chatterley_long-lasting_2019} and molecular complexes~\cite{yang_helium_2012,schouder_structure_2019}. This includes for instance homo-dimers of metal atoms~\cite{claas_wave_2006,lackner_spectroscopy_2013,thaler_long-lived_2020}, where neither IR nor microwave spectroscopy can be used.

We believe the rotational dynamics reported here is a consequence of superfluidity of the helium droplets. According to the power spectra, shown in \autoref{trace-fft}, the maximum $J$ and thus rotational energy, $E_\text{rot}$(~\autoref{eq:rotational-energy}), are $J$=5,~1.9 cm$^{-1}$ for \ce{OCS}; $J$=10,~2.2 cm$^{-1}$ for \ce{CS_2}; and $J$=9,~1.3 cm$^{-1}$  for \ce{I_2}. In all three cases, $E_\text{rot}$ is below the roton energy. Thus, the free molecular rotation we observe is consistent with the weak coupling between molecular rotation and phonons in this low-energy regime~\cite{choi_infrared_2006}. If stronger alignment pulses are applied, higher rotational states will be excited. Then the free-rotor description is no longer expected to be valid since the coupling to the phonons (rotons) strongly increases and the high density of states above the roton energy leads to fast decay~\cite{zillich_roton-rotation_2004}. Furthermore, the rotational level structure will not continue to be given by \autoref{eq:rotational-energy}, which predicts that E$_\text{rot}$ starts decreasing when $J$ exceeds a value, $J_{th}$, e.g. for \ce{OCS} $J_{th}$~=~10. Time-resolved rotational dynamics measurements may provide an understanding of this unexplored regime of rotational states.

Finally, previous studies have shown that an intense fs laser pulse can deposit so much rotational energy in a molecule that it transiently decouples from its solvation shell~\cite{shepperson_laser-induced_2017}. Such a process will bring the molecule-He system far away from equilibrium. Measuring the rotational dynamics for different delays between the distortion and a subsequent weak alignment pulse will allow to explore how long it takes to restore the equilibrium, gauged by when the alignment trace becomes identical to a reference trace recorded without the distortion pulse. Alternatively, the rotational echo technique may enable real-time characterization of the molecule-helium droplet coupling through measurement of the time constant for coherence loss of rotational wave packets~\cite{karras_orientation_2015,rosenberg_echo_2018,zhang_rotational_2019}.

H.S~acknowledges support from the European Research Council-AdG (Project No. 320459, DropletControl) and from The Villum Foundation through a Villum Investigator grant no. 25886. M.L.~acknowledges support by the Austrian Science Fund (FWF), under project No.~P29902-N27, and by the European Research Council (ERC) Starting Grant No.~801770 (ANGULON). G.B.~acknowledges support from the Austrian Science Fund (FWF), under project No.~M2461-N27. I.C.~acknowledges the support by the European Union's Horizon 2020 research and innovation programme under the Marie Sk\l{}odowska-Curie Grant Agreement No.~665385. Computational resources for the PIMC simulations were provided by the division for scientific computing at the Johannes Kepler University.


%


\end{document}